\documentclass[runningheads]{llncs}
\usepackage[height=235mm,width=155mm,center]{crop}

\usepackage{orcidlink}
\usepackage[T1]{fontenc}
\usepackage{graphicx}
\usepackage{hyperref}
\usepackage{color}

\usepackage{xcolor}
\usepackage{booktabs}
\usepackage{multirow}
\usepackage{subcaption}
\usepackage{fancyhdr}

\colorlet{punct}{red!60!black}
\definecolor{background}{HTML}{EEEEEE}
\definecolor{delim}{RGB}{20,105,176}
\colorlet{numb}{magenta!60!black}

\fancypagestyle{FirstPage}{
\lfoot{*Author version. The final publication is available at Springer via \url{http://dx.doi.org/10.1007/978-3-031-57978-3_12}.} 
}

\begin{document}
\title{Secure and Privacy-Preserving Authentication for Data Subject Rights Enforcement\textsuperscript{*}}
\titlerunning{Authentication for Data Subject Rights Enforcement}
%
\author{Malte Hansen\orcidlink{0000-0003-4622-3819} \and
Andre Büttner\orcidlink{0000-0002-0138-366X} }

\authorrunning{M. Hansen, A. Büttner}
%
\institute{Department of Informatics, University of Oslo, Norway\\
\email{\{maltehan,andrbut\}@ifi.uio.no}
}
\maketitle              

\begin{abstract}
In light of the GDPR, data controllers (DC) need to allow data subjects (DS) to exercise certain data subject rights. A key requirement here is that DCs can reliably authenticate a DS. Due to a lack of clear technical specifications, this has been realized in different ways, such as by requesting copies of ID documents or by email address verification. However, previous research has shown that this is associated with various security and privacy risks and that identifying DSs can be a non-trivial task.
In this paper, we review different authentication schemes and propose an architecture that enables DCs to authenticate DSs with the help of independent Identity Providers in a secure and privacy-preserving manner by utilizing attribute-based credentials and eIDs. Our work contributes to a more standardized and privacy-preserving way of authenticating DSs, which will benefit both DCs and DSs.

\end{abstract}

\section{Introduction}\thispagestyle{FirstPage}
Since the GDPR entered into force in May 2018 \cite{gdpr}, citizens of the EU have certain data subject rights (DSR). Therefore, the data controller (DC) of any service that collects personal data is required to implement measures for letting the rightful owner of the data, the data subject (DS), exercise their DSRs. For this, a DC must reliably identify and authenticate the data owner to ensure that only legitimate DSs can access their data. However, the GDPR does not specify concrete, technical, or non-technical measures to do this. Consequently, there are currently different implementations of DS authentication used in practice. On most popular online services such as Google or Meta, users can access their data and exercise their DSRs by authenticating to their user accounts. Yet, there are many other use cases where user data is collected, but no user accounts exist. It has, therefore, become common practice to authenticate DSs by requesting a copy of their ID document to match the information with a data set or by verifying, e.g., an email address. However, these methods can be problematic as they might put the DS at risk and are often not compliant with the GDPR. For instance, in 2022, a media company in the Netherlands was fined for unnecessarily requesting full ID documents, forcing them to change their procedure to email verification \cite{edbp2022dutch}.

The establishment of a digital identity is another vital endeavor driven by the European Commission to provide EU citizens with a unified way to authenticate to digital services, even across countries, in order to push forward digitalization. For that purpose, the first European regulation on electronic identification, authentication, and trust services (eIDAS) was proposed in 2014, specifying requirements for implementing eIDs~\cite{EUDigitalIdentity2014}. Due to a reluctant adoption of eIDs by different countries and new requirements by different stakeholders, the regulation was updated in 2021 in that eIDs shall be based on the EU Digital Identity Wallet~\cite{EUDigitalIdentity}, following the paradigm of the self-sovereign identity (SSI)~\cite{preukschat2021self}. With this, European citizens will be able to manage their own attributes, such as personal information otherwise found on an ID document or other official documents like their driver's license on a digital wallet on their phone.

We identified that eIDs can also be helpful in the context of DSR enforcement. The fact that citizens can present only specific attributes that are issued by a trusted authority and are necessary to verify a DS's identity makes it a useful way of authentication. However, this requires a flexible architecture that can be used by DCs with relatively low effort for it to be practically feasible. In this paper, we describe in detail how DS authentication based on eIDs can be implemented. Furthermore, we discuss different scenarios to showcase how our approach can be used in practice. The main contributions of this paper can thereby be summarised as follows:
\begin{itemize}
    \item A concept for using eIDs and SSI in the context of DSR. 
    \item An architecture for secure and privacy-preserving authentication of DSR requests with eIDs.
    \item An analysis of the architecture in consideration of the EU Data Strategy.
\end{itemize}
The content of this paper is structured as follows. Section \ref{sec:background} gives detailed background information on DSRs and eIDs. In Section \ref{sec:relatedWork}, the related work is summarised. After that, Section \ref{sec:reviewAuthenticationSystems} provides an overview of the different authentication models and discusses eID as a solution for authentication in the context of DSRs. Section \ref{sec:proposedSolution} describes the architecture of our solution as well as example scenarios and an analysis. In Section \ref{sec:discussion}, the proposed solution is discussed. Section \ref{sec:Conclusion} summarizes our results and gives an outlook on future work. 

\section{Background}\label{sec:background}
The GDPR~\cite{gdpr} grants several DSRs to European citizens. As an example, with the Right to Erasure, a DS can request the deletion of all their information by a DC, while the Right of Access (RoA) allows a DS to request a copy of all their data from a DC. The disclosure of such a RoA result holds great risks. Hence, secure authentication is a core element of the enforcement process for many DSRs.

In fact, DSR enforcement was shown to be prone to authentication attacks. For RoA requests, several studies have gained unauthorized access via the utilization of social engineering attacks, leveraging the requirement for human actions in the authorization process~\cite{cagnazzo2019gdpirated,di2019personal,pavur2019gdparrrrr}. In an eID scenario, these weaknesses extend to possible abuse of DSRs by an authoritative government ~\cite{lauradoux2022can}. To get rid of these weaknesses, the need for an alternative authentication process for DSRs has been identified, especially in scenarios where the DC does not have any established authentication method~\cite{hansen2022generic}.

This demand for secure DSR authentication becomes even greater in the face of the European data strategy~\cite{EUDataStrategy}. An essential part of the strategy is the establishment of a single European data market, fostering data-sharing inside the EU. As a result of this strategy, the Data Act~\cite{da} facilitates data-sharing across different sectors. Further, the Data Governance Act~\cite{dga} introduces Data Intermediaries (DI), acting as a sort of data broker between different DCs and DSs while fostering DSRs. These developments will lead to increased data flows between more actors. Consequently, the existence of secure and reliable DSRs gains more importance. Another important development regarding authentication in the European data strategy is the relevance of eIDs. Building upon eIDAS, a framework for interoperability of eIDs inside Europe, the European Commission has proposed the Digital Identity Regulation~\cite{EUDigitalIdentity}, looking to establish a European Digital Identity available for any citizen, resident, or business inside the EU.

An authentication method that has recently gained a lot of traction is attribute-based credentials (ABC)~\cite{sabouri2015abc4trust}, with ENISA identifying privacy-enhanced ABCs as an important technology in the European data strategy~\cite{Enisa2023Engineering}. While this solution is most practical in use cases where a DS only has to prove one specific attribute, such as its age or citizenship, it has to be considered if and how ABCs can be utilized in an authentication process in a DSR enforcement scenario.

\section{Related Work}\label{sec:relatedWork}
Previous research has analyzed how DSRs can be performed in practice. For instance, Boniface et al.~\cite{boniface2019security} found that in several cases where Service Providers (SP) do not offer user accounts, DSs had to send in a copy of their ID document to verify their identity and sometimes even only to check on the eligibility of RoA. This, however, contradicts the principle of data minimization and might allow a malicious actor to misuse the document for impersonation. They also highlight the challenge for third-party trackers to re-identify a DS and propose a cookie with a pseudonym derived from a DS's email address. Another approach is the concept of Data Subject Rights as a Service (DSRaaS)~\cite{DSRaaS}. DSRaaS defines Identity Providers~(IdP) for DSRs, e.g., embedded in the responsibilities of a DI, that act as IdPs for DCs lacking the resources to authenticate a DS themselves properly. 

Similar studies observed that beyond ID documents, email addresses and phone numbers are often used by DCs to authenticate DSs. It was discovered that there are several ways for a malicious actor to access the data of another DS, e.g., by forging an ID document using open-source intelligence techniques or by requesting the data from a non-corresponding email address~\cite{di2019personal,di2022revisiting}. Furthermore, in the study by Urban et al.~\cite{urban2019study}, many companies did not respond to data requests within the prescribed period of 30 days or at all.

Other work focused on identifying a DS within a data set. It might be challenging to create a reliable mapping between DSs and the data that actually falls within their DSR, as it highly depends on what identifiers are present and how reliably their ownership can be verified. Therefore, different levels of identification are defined and discussed~\cite{purtova2022knowing}.

Most of the prior research on eID addresses the implementation challenges and compliance in different countries like the UK~\cite{tsakalakis2016identity} or Estonia and the Netherlands~\cite{lips2020eidas} based on the first eIDAS proposal~\cite{EUDigitalIdentity2014}. Furthermore, some work looked at the security of eIDAS implementations. For example, a security analysis on different eID implementations was conducted that discovered that many implementations were vulnerable to XML-based attacks~\cite{engelbertz2018security}. Another survey conducted in 2021~\cite{sharif2022eidas} studied what authentication protocols and methods were used by the eID implementations of different countries, revealing that several countries only provided a low level of assurance, thus providing low security properties. Beyond that, much research on eIDAS has been done in the context of universities~\cite{gerakos2017electronic,berbecaru2019electronic,alonso2020enhancing}. 

The research above discloses security and privacy issues of current DSR authentication deployments. Yet, to the best of our knowledge, there is no proposal for a standardized way to overcome these problems. Our work contributes by investigating a new use case for eIDs that addresses the challenges of identifying and authenticating DSs, as described in the work mentioned above. In particular, we propose a solution that mitigates the risk of impersonation and helps to minimize the data in a standardized manner.

\section{Review of Authentication Systems}\label{sec:reviewAuthenticationSystems}
Authentication systems differ in who or what is controlling the identities of users. 
Throughout the years, authentication systems have been implemented in various ways. The main distinction lies in the different models, which are \textit{centralized identity management} (CIM), \textit{federated identity management} (FIM), and \textit{decentralized identity management} (DIM) \cite{avellaneda2019decentralized}. The latest\footnote{At the time of writing} proposal for the European eID is based on the latter model that aims to provide European citizens with a secure and privacy-preserving authentication solution for online services.

\subsection{Identity Models}\label{sec:models}

\begin{table}[ht!]
    \centering
    \caption{Summary of advantages and disadvantages of the different identity models. Note that this list is non-exhaustive.}
    \label{tab:summary_identity_models}
    \begin{tabular}{p{2.1cm}@{\hskip 0.2cm}|@{\hskip 0.2cm}p{4.6cm}@{\hskip 0.2cm}|@{\hskip 0.2cm}p{4.6cm}}
    \toprule
        \textbf{Model}  & \textbf{Advantages} & \textbf{Disadvantages}   \\
    \midrule
        Centralized & \textbullet \hspace{0.1cm}Service can be independent of a third-party  \newline
        \textbullet \hspace{0.1cm}No tracking of users & \textbullet \hspace{0.1cm}Higher implementation and user management effort by the service \newline \textbullet \hspace{0.1cm}Users need to keep track of multiple credentials \\\midrule
        Federated & \textbullet \hspace{0.1cm}Lower implementation and user management effort by the service \newline \textbullet \hspace{0.1cm}Less credentials for the users to keep track of
        & \textbullet \hspace{0.1cm}Service depends on third-party IdP \newline \textbullet \hspace{0.1cm}Users may have low control about which attributes are shared \newline 
 \textbullet \hspace{0.1cm}IdP can track user activity  \\\midrule
        Decentralized & 
        \textbullet \hspace{0.1cm}Service can be independent of a third-party \newline \textbullet \hspace{0.1cm}Control over attributes remains with the users \newline \textbullet \hspace{0.1cm}Lower implementation and user management effort  by the service & \textbullet \hspace{0.1cm}Users need to take care of backups  \\
    \bottomrule
    \end{tabular}
\end{table}

\begin{figure}[t]
\centering
\begin{minipage}{.5\textwidth}
  \centering
  \includegraphics[width=.95\linewidth]{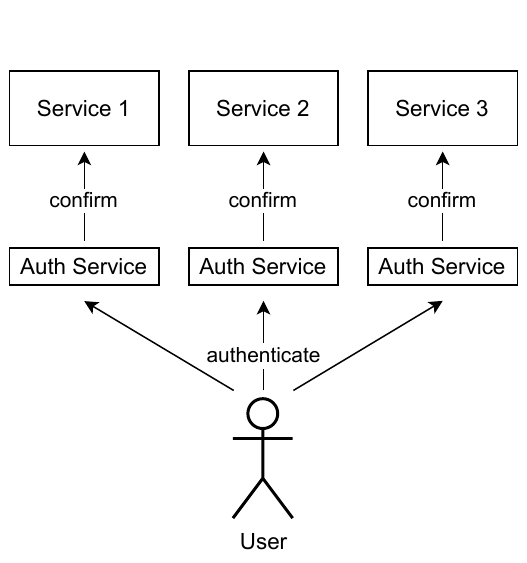}
  \caption{Centralized identity model}\label{fig:cim}
\end{minipage}%
\begin{minipage}{.5\textwidth}
  \centering
  \includegraphics[width=.95\linewidth]{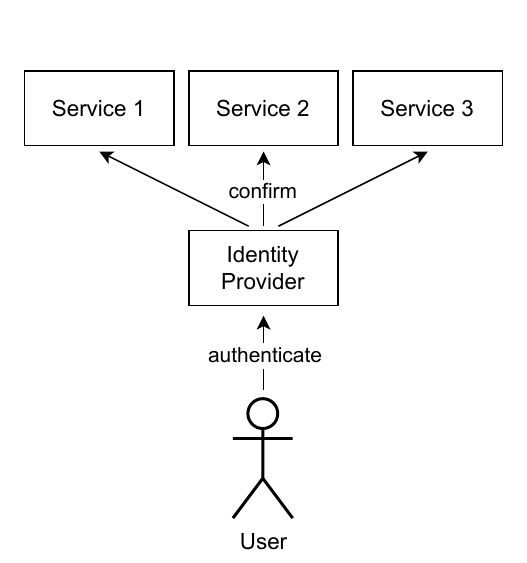}
  \caption{Federated identity model}
                \label{fig:fim}
\end{minipage}
\end{figure}

\begin{figure}
    \centering
    \includegraphics[width=\textwidth]{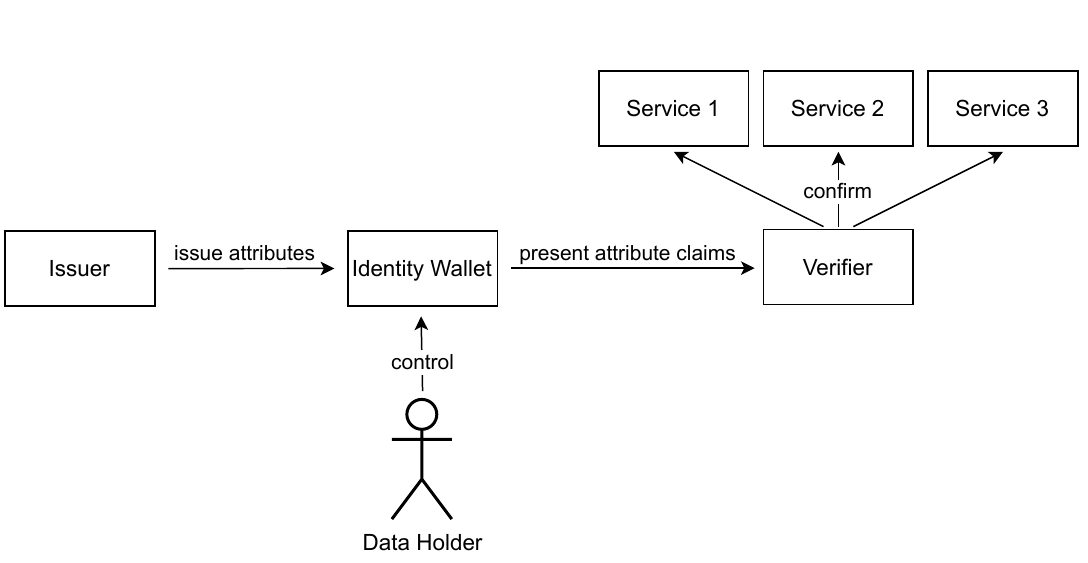}
    \caption{Decentralized identity model based on ABCs}
    \label{fig:dim}
\end{figure}

The different identity models come with several advantages and disadvantages. Table~\ref{tab:summary_identity_models} provides a brief overview of these. 
In a CIM system, users have accounts specifically for a service or within one enterprise, as shown in Fig.~\ref{fig:cim}. This makes the respective digital identity independent from any other service or outside an enterprise. Compared to a federated system, this is more privacy-friendly as users can not easily be tracked across different services. However, it is less convenient for both the SP and the user. The SP needs to implement an authentication service and handle the user database. A user has to manage many different credentials for the distinct services, which often leads to the choice of less secure passwords \cite{gaw2006password}.
Regarding DSRs, if a service with CIM provides sufficiently secure authentication methods such as multi-factor authentication, it can authenticate a DS for DSR requests without requiring additional data, thus not being affected by the risks described in Section \ref{sec:relatedWork}. 

In FIM systems, the service, usually referred to as relying party (RP), outsources the identity management to an IdP. Users must authenticate against this IdP and grant access to certain account information to sign in to this RP. This is illustrated in Fig.~\ref{fig:fim}. Depending on the IdP, a user might have limited control over how much information is shared between RP and IdP. Furthermore, the IdP can track the services accessed by a user, significantly impacting the user's privacy. However, with an FIM, the RP's implementation effort is relatively low since the IdP covers user management and authentication. The RP only needs to have a trusting relationship with the IdP and serve as a client. Commonly, this is implemented with widely-known, standardized protocols such as OAuth2 \cite{rfc6749}, OpenID Connect \cite{OpenIDConnect} or SAML \cite{oasis2005saml}. Moreover, a user has fewer credentials to handle, as the same credentials are used for multiple services with the same IdP.

DIM systems aim to let users keep control of their identity attributes instead of IdPs. For that purpose, there are \textit{attribute-based credentials} (ABCs)~\cite{sabouri2015abc4trust}, which are a step towards a more user-centric approach. Using an architecture consisting of an \textit{issuer}, \textit{verifier}, \textit{data holder}, it provides important privacy properties such as unlinkability and data minimization. The issuer is responsible for issuing valid attributes to the legitimate data holder who stores these credentials in their identity wallet. The data holder can then send attribute claims to a service that can verify these claims or use a third-party verification service to confirm the validity of the presented attributes. This is shown in Fig.~\ref{fig:dim}.
The paradigm behind this is referred to as \textit{self-sovereign identity} (SSI) \cite{preukschat2021self}, which describes that users have complete control over their own attributes. It also often involves using decentralized architectures like distributed ledger technology (DLT) for public verifiability \cite{muhle2018survey}. Since this is relatively new, users might have to get used to handling an identity wallet. Few studies on digital identity wallets have shown that users manage to use them to a certain degree but have problems creating and using a backup \cite{khayretdinova2022conducting,satybaldy2022usability}. However, one can assume that attributes can be re-issued in many cases. Also, it can be more convenient for users to have all their attributes and credentials in one place, which can be beneficial in the long run. In terms of security, it is crucial that the identity wallet is implemented with sufficient protection measures, e.g., using secure hardware and access control to prevent the credentials from being stolen. From an SP perspective, it can have similar benefits as for FIMs, as it is possible to combine it with FIM protocols like OpenID Connect to outsource the verification of credentials and thus does not require the implementation of any further authentication methods.

Importantly, SSI is considered in the latest eIDAS regulation~\cite{EUDigitalIdentity} for the EU-wide implementation of eIDs. After the slow adoption of the first eIDAS regulation in 2014, based on a FIM~\cite{EUDigitalIdentity2014}, there has been an updated proposal (sometimes referred to as eIDAS 2.0). The new eID is based on SSI and builds on the EU Digital Identity (EUDI) Wallet.

\subsection{EU Digital Identity Wallet}\label{sec:eudiw}
According to the Digital Identity Regulation, ``at least 80\% of citizens should be able to use a digital ID solution to access key public services by 2030'' \cite{EUDigitalIdentity}. Therefore, EU countries must ensure that an eID scheme that is compliant with this regulation is provided by 2030. A key requirement, already since the first eIDAS proposal, is to provide cross-border interoperability. In addition to the regulation, an architecture and reference framework has been created~\cite{EUDIWalletARF_2023} to provide technical guidelines. The core element in the specified architecture is the EUDI Wallet instance, which is controlled by the user. Such a wallet can be either an application running locally on a mobile device or remotely as an online service. While the former case allows a user to protect their wallet physically, strong access control measures like MFA are required in the latter case to protect the wallet from being compromised.
An identity wallet is meant to store different types of attributes, such as personal identifiable data (PID), which are attested and issued by trusted providers. These attributes can then be presented as verifiable claims to a relying party, which can verify the authenticity of these claims.

The reference framework also specifies which personal attributes are considered in the eIDAS framework. These attributes are basically those that can be found in ID documents. While the first name, last name, date of birth, and unique identifier are mandatory, other attributes are optional. These attributes are issued by specific PID providers, which must follow the issuing requirements for PID.
Other types of attributes, e.g., driving licenses or digital payments, are issued by so-called (Non-)Qualified Electronic Attestation of Attributes ((Q)EAA) Providers \cite{EUDIWalletARF_2023}.

\subsection{Using eID for Authenticating DSR Requests}

The European Commission considers various use cases for their pilot implementation of the EUDI Wallet, such as access to electronic government services, opening a bank account, claiming prescriptions, and many more~\cite{EUDIWalletImpl_2023}. The use of eID in connection with DSRs has been suggested \cite{hansen2022generic}, but to the best of our knowledge, there has not yet been proposed a concrete concept for this.

As shown in Section \ref{sec:relatedWork}, prior research has revealed that the state-of-the-art for authenticating DSs for RoA requests has many issues concerning their security and privacy.
We aim for an approach that neither suffers from weak authentication methods, such as passwords or email address verification nor violates data minimization, e.g., by requesting a copy of their complete ID document.
Especially since the eIDAS regulation provides an EU-wide approach to authenticate citizens, this might be a particularly relevant authentication solution. The eIDs can help enforce DSRs in a more standardized, secure, and privacy-preserving manner.

We consider typical scenarios where services do not provide means of authentication, for instance, when using online retail as a guest user or advertising companies that collect information by cookies. In such cases, DCs need to put in place alternative methods to identify a DS in order to allow them to access their data. With the infrastructure of the EUDI Wallet, users can use the attributes stored in their identity wallet to present identity claims that overlap with the data held by the DC. This allows for matching between the data sets and their rightful owner. If a DC can verify the attributes presented by a DS, he can let them exercise their DSRs.

The required attribute claims will vary depending on the data sets. In many cases, a DC can request verifiable claims for the types of PID attributes that are included in the data set, e.g., first name, last name, and address.  However, it is crucial that a DS can be identified reliably and that a sufficient number of attributes is verified. For example, verifying a DS's address is likely not sufficient. In cases where a data set is not sensitive, it might be acceptable to have lower requirements for the uniqueness of the requested attribute claims. For that purpose, the levels of identification discussed in \cite{purtova2022knowing} can be useful in deciding about such requirements. With eIDs, DCs have no reason to request complete ID documents from users. This mitigates the risk of their misuse and prevents the disclosure of irrelevant yet sensitive information. In addition, authentic ID information can only be obtained from a reliable and most likely governmental provider, making forgery significantly more difficult. 

In cases where other means are used to authenticate a DS, such as email address or phone number verification, DCs should consider verifying other attributes that are included in the data set. Depending on the concrete use case, there might be verifiable attributes like the name, payment transactions, or other technical features. It is also conceivable that email providers themselves issue an attribute to verify the ownership of an email address, which might mitigate the risk of improper email validation. However, in that case, email providers still need to ensure secure authentication.

\section{Attribute-based eID Authentication for Data Subject Rights Enforcement}\label{sec:proposedSolution}
As stated in Table \ref{tab:summary_identity_models}, one of the advantages of a DIM is the control the user has over its credentials. This aligns with the DSRs' purpose of giving DSs control over their data. However, a pan-European solution also requires a uniform solution accessible to all DSs and DCs. Further, data minimization is a major concern during authentication. To meet these criteria, the proposed architecture combines an eID authentication framework with ABCs to provide a secure authentication scheme for DSR enforcement while reducing the amount of data disclosed during the process.

\subsection{Architecture}
\begin{figure}[t]
    \centering
    \includegraphics[width=.8\textwidth]{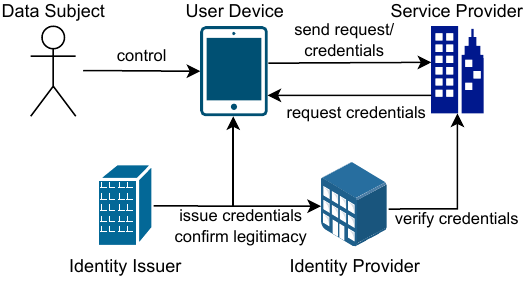}
    \caption{Overview of Components in the Architecture}
    \label{fig:components}
\end{figure}

Inspired by Papadamou et al.'s privacy-preserving architecture for device-centric and attribute-based authentication~\cite{papadamou2019killing}, we introduce an architecture focused on assigning roles according to the European data strategy (see Fig.~\ref{fig:components}). 
It consists of a \textbf{(1) User Device}, an eID that can either be an RFID chip on a physical card or a standalone application on a device, such as the DSs smartphone. It is also possible for the application to be outsourced to a cloud. The User Device holds the credentials of the eID. A \textbf{(2) Service Provider} is a DC requiring authentication of a User Device, while a \textbf{(3) Identity Provider} is a trusted party, like a DI, receiving authentication requests for a User Device from the SP. For these requests, an ABC authentication scheme is used. Valid credentials depend on the relevant data sets for the DSR request, as well as the catalog of credentials included in the implementation of the eID. The IdP can either verify the credentials sent from the User Device to the SP or verify credentials for a specific DS itself. The \textbf{(4) Identity Issuer} is a neutral, third entity issuing the User Device's instance of the eID. For this purpose, they exchange information with the governmental entity that creates the national ID. Outsourcing the role of the Identity Issuer from the issuing government entity is recommended to combat the attack vector of a malicious government or entity. The Identity Issuer also maintains the instance of the User Device by performing updates or recoveries. Additionally, they confirm the expiration date of the eID to IdPs and the User Device. Additionally, they can possibly negotiate valid credentials. A DS might want to add or remove some attributes from the catalog of valid credentials or restrict certain attributes to specific use cases. While the Identity Issuer has an independent role in the architecture, they can also function as an IdP at the same time.

\subsection{Process Flow}
Applying SSI to the architectural components leads to a process where the User Device holds all the credentials and the IdP only verifies them to the SP. However, an FIM solution may be preferred in specific circumstances, as will be discussed in Section \ref{sec:analysis}. In the FIM approach the process is mainly done between the SP and IdP. For both approaches, we assume that the Identity Issuer has issued a valid eID with a sufficient expiration date to the User Device and confirmed this to the IdP. 

\begin{figure}[t]
    \centering
    \includegraphics[width=\textwidth]{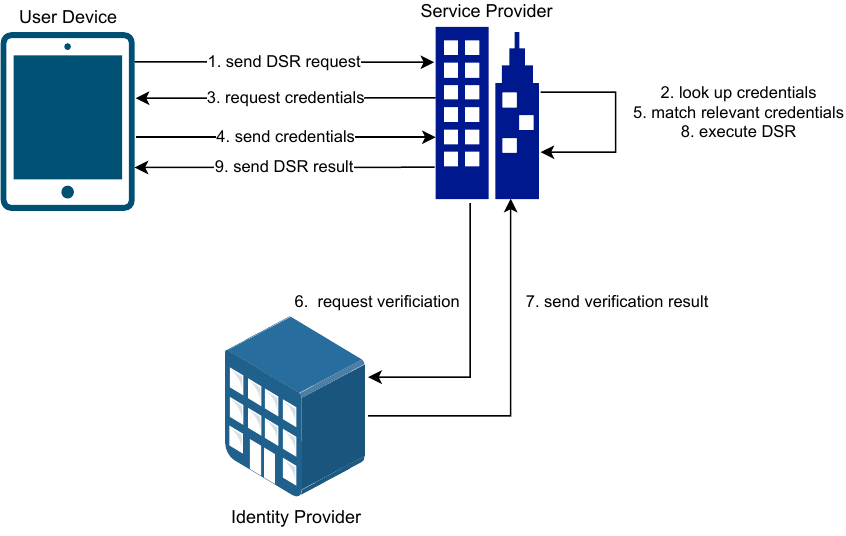}
    \caption{Simplified flow of an SSI authentication process}
    \label{fig:process_flow}
\end{figure}

The SSI approach, see Fig.~\ref{fig:process_flow}, starts with the DSR request by a DS for a certain data set held by a DC. So, the User Device sends a request with a payload to the SP. The payload includes information on which DSR shall be invoked and which data sets are subject to this request. This can be any single specific data set, all data sets with information about the DS the SP holds, or anything in between. The SP looks at which valid credentials can exist in their data sets, e.g., by taking the customer number for which the request was issued and querying it over the catalog of valid credentials. The SP now sends a request for these credentials to the User Device. To prevent possible attack scenarios, this request is the same for any User Device contacting the SP. To see which credentials are relevant, the SP looks up which attributes exist in the corresponding data sets. After matching the stored credentials with the ones received in the response from the User Device, they are forwarded to the IdP, who verifies them. The following confirmation or decline of the response from the IdP does not contain any information about which credentials are correct or wrong. Finally, in case of successful verification, the DSR can be processed by the SP. 

\begin{figure}[t]
    \centering
    \includegraphics[width=\textwidth]{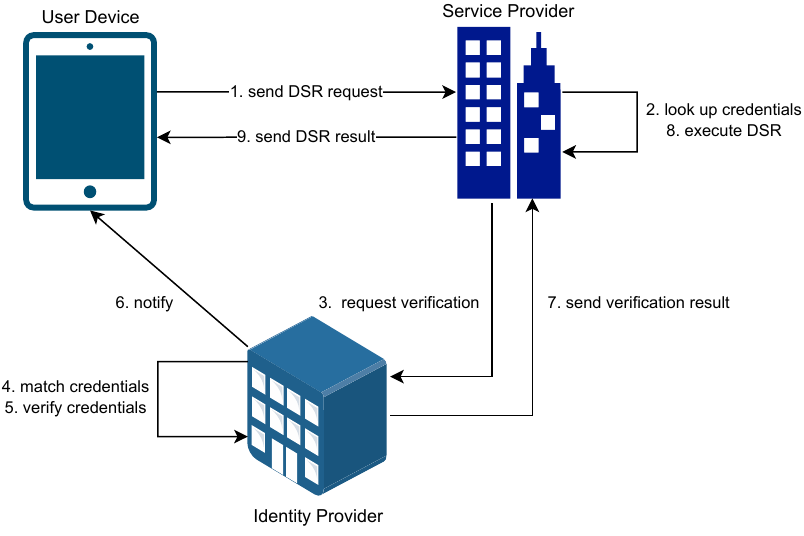}
    \caption{Simplified flow of a FIM authentication process}
    \label{fig:process_flow_ip}
\end{figure}

In the FIM approach, see Fig.~\ref{fig:process_flow_ip}, the request is initiated in the same way. The SP again looks for valid credentials in the relevant data sets. The SP then sends a request for authentication containing the required credentials to the IdP, which is doing the mapping in this scenario. The IdP then confirms the credentials. To prevent misuse of the authentication system by a malicious SP, the IdP notifies the User Device about the exchange with the SP. Again, the response to the SP does not contain any information about which attributes are correct or wrong. Alternatively, the DS could initiate the request in this approach over the IdP. The User Device would then send the parameters for the request to the IdP, which would forward them to the SP.

\subsection{Analysis}\label{sec:analysis}
The main application of this model is for an attribute-based authentication of DSR requests. It allows the DC to determine a threshold of credentials to authenticate a DS while simultaneously verifying these credentials via a neutral, certified party. How to reach a sufficient threshold for an unambiguous authentication is highly complex and depends on the context and credentials at hand, therefore requiring further research. However, the architecture allows a variety of attributes to be included in this process. For example, the address and date of birth can be taken from the eID and mapped to an order number. This means that to authenticate a data set, it is only required to reveal a certain number of attributes instead of the entire catalog of credentials, as is often the case in a traditional authentication scenario. A precondition for this way of authentication is the existence of the credentials available in the eID solution in the data sets. Therefore, options for a meaningful expansion of these attributes, such as the (Q)EAA Providers in the EUDI Wallet, must be developed further. Using the architecture, a DS can also prove that a data set belongs to it based on specific credentials, even in cases where the data sets are not linked beforehand. This can also be utilized in scenarios where data sets belong to multiple DSs, e.g., in a shared streaming account. Since you do not need to reveal the user ID or name in the request, the DC can not learn which DS issued the request. This enables the possibility of an 'anonymous data subject right request'.

The process of searching the data sets for the credentials is greatly facilitated by a universal data model, as introduced in previous work~\cite{datamodel}. By attaching metadata about the category of personal, such as a name, address, or date of birth, to the data sets, you can use these keywords to build your query.

A potential issue is a mismatched authentication due to incorrectly stored or guessed attributes or an impersonation attack based on credentials similar to the attacked DS. While the risk can not be completely mitigated, the IdP can see which eID issued the request. Hence, they can initiate adequate steps to address this problem as long as the eID itself is not compromised. 

Besides their general characteristics, described in Section~\ref{sec:models}, the two different approaches introduced come with their own advantages and disadvantages. In the SSI approach, the SP always requests all credentials that can theoretically exist in their data sets. This is done so an attacker cannot learn that a particular data set includes a specific attribute by instigating an intentionally wrong request. While the values can be hidden by, for example, matching only hashes instead of clear text, this nonetheless means that the SP could learn about the existence of a credential not included in the data set. For the FIM approach, a similar issue exists, as a malicious IdP can make the same attack based on the reduced catalog of only relevant credentials. With strong certification and auditing for the IdPs, this is easier to mitigate though. The main difference between the two models comes in the distribution of competencies. The SSI flow gives most of the processing work to the SP, including the determination of the authentication threshold. Consequently, the IdP, as it acts as only a verifier, requires less implementation and acquires less information about the DS. Compared to a traditional approach, this still facilitates the implementation for the SP, as it can rely on the eID to verify the DS's claims. The FIM approach, on the other hand, puts a lot of responsibility on the IdP. This demands strong compliance control of the IdP. It is, however, advantageous in scenarios where the SP does not have the know-how or resources available to determine the authentication threshold reliably. For SPs that are hard to audit and assert compliance with, as may be the case with non-European DCs, or the DS does not trust the SP, this is also a preferred solution. Additionally, the FIM approach could also be realized for DSs without an eID. As the use of an EUDI Wallet by citizens is not mandatory under the legislative proposal and the distribution of them has only started, this is an important factor to consider during this transition period.

Considering the direction the EU Data Strategy~\cite{EUDataStrategy} is heading, the proposed architecture fills an important gap in ensuring secure and reliable authentication for DSRs. With the introduction of data spaces and the general increase of data sharing, it will become more prevalent that a DC will collect or receive personal data that is not linked to an existing authentication solution or other information, like e-mail addresses, that could be used to authenticate a DS easily. The DI, already at the center of this new data landscape, is a natural fit to fulfill the responsibilities of the IdP. Especially as one of the obligations of DIs is to facilitate DSR enforcement. For this purpose, the DI can serve the role of the Identity Issuer as well. Additionally, the involvement of the IdP and Issuer would mesh well with concepts like DSRaaS, where the DSRaaS Provider, or DI, already has a lot of competencies and responsibilities.

\section{Discussion and Open Issues}\label{sec:discussion}
\subsubsection{Authentication threshold}
As mentioned in Section \ref{sec:analysis}, it is important to determine a sufficient, at best generalized, threshold for a combination of credentials to identify a DS. However, this is a very challenging task as the threshold should be neither too high nor too low, and each data set might require a different set of attributes to achieve unambiguous identification. The threshold can also be impacted by the category of personal data in question. Sensitive data requires additional protection~\cite[Art. 9]{gdpr} and thus demands a stronger threshold. Further, the addition of new information to a context can always change this threshold. While privacy-enhancing technologies can be used to strengthen anonymization, they do not necessarily protect against re-identification through new information at a later point in time.
\subsubsection{Credential negotiation}
A possibility to strengthen the authentication scheme is the addition of more credentials, which the EUDI Wallet addresses with the (Q)EAA Providers. This could possibly also be extended toward technical identifiers. However, technical identifiers are not necessarily unique and will likely change over time, e.g., shared devices and IP addresses in public spaces. This makes them more unreliable. While a combination of these factors might reach a satisfactory threshold to identify a DS, regular exchange of information about these factors with the DS would be required to keep them up to date. Considering the privacy risks and contradiction of data minimization as well, it is highly questionable whether such an approach would improve the authentication process. 
Another possibility is the usage of derived attributes by the SP, such as the estimated address, gender, or age. This does hold potential, as an estimated age range or postal code can be used as a weak credential to reach a sufficient authentication threshold. However, it does include risks as well as the attributes might be derived incorrectly or based on outdated information. In the case of a rough estimate like an age range, it might also allow for too much leeway and lead to verifying a different person. Consequently, derived attributes should be labeled as those and only be used as a supplement to other credentials where only additional weak factors are required to reach a sufficient authentication threshold. Additionally, the accuracy of the underlying algorithm must be high enough so the usefulness of the additional credentials outweighs the risks it contains.
\subsubsection{Semantics}
The semantics of data during the verification process is another issue that has to be considered during implementation. Lower- and upper-case writing, as well as localization of values with language-specific symbols, can lead to mistakes that would have to be specifically addressed by a process or manual intervention. Ideally, this should already be considered during the data creation by enforcing certain standardized, structured data formats. Already existing data sets will need to be normalized for that purpose.
\subsubsection{Non-European Data Subjects}
Another question is how non-European residents can use the authentication architecture, as they do not possess a European eID. While it is possible to create a service to transform non-EU eIDs, it is not guaranteed that these residents even own an eID from their country. Therefore, a service to give an eID to non-EU is required anyway. This could be integrated into migration offices or the IdPs introduced here accordingly.

\section{Conclusion}\label{sec:Conclusion}
This paper addresses the problem of identifying and authenticating data subjects in order to let them exercise their DSRs securely, reliably, and with respect to their privacy. To this end, various authentication schemes are first reviewed, and finally, an architecture based on eID and ABCs is proposed to provide a standardized authentication scheme. It includes User Devices belonging to the DSs, SPs, addressing the DCs, and IdPs and Issuers, which can find a meaningful representation in the European Data Strategy in the form of the DIs. The scheme introduces two different approaches, namely an SSI and an FIM approach, to meet different demands. This architecture can be used in particular by data controllers that do not offer their own authentication scheme, request copies of ID documents, or authenticate users by somewhat insecure methods like email address verification. Consequently, this architecture allows both DSs and DCs to rely on an eID solution for a secure authentication that only reveals the necessary attributes of the DS instead of its full identity.

In future work, an implementation of this model should be tested and improved. Furthermore, as the realization of the proposed architecture highly depends on the EU-wide adoption of eIDAS, further studies on the implementation of eIDAS should be conducted. While the architecture considers scenarios where eIDAS is not usable, this process must be more streamlined to not unnecessarily bloat the architecture. An open issue that must be solved before this scheme can be employed is the reliable construction of a sufficient authentication threshold that is both functional and secure. Finally, it has to be investigated further how this architecture can be used in conjunction with non-European DSs and DCs.
%
%
%
 \bibliographystyle{splncs04}
 \bibliography{paper}
\end{document}